\pdfoutput=1
\documentclass[aps,prd,onecolumn,nofootinbib,preprintnumbers]{revtex4}
\usepackage{amssymb}
\usepackage{graphicx}
\usepackage{latexsym,amssymb, amsmath,color}

\newcommand{\func}[1]{#1}

\begin{document}
\preprint{MCTP-12-30}
\title{Review and Update of the Compactified M/string Theory Prediction of the Higgs Boson
Mass and Properties \footnote{Invited review for the International Journal of Modern Physics A}$\hspace{2mm}$\footnote{This is a review and update of the August 2011 prediction of the
Higgs boson mass and properties, based on invited plenary talks at
the Beijing Supersymmetry 2012 conference, the International String Phenomenology 2012 annual
meeting (University of Cambridge), the Newton Institute Programme on Strings and M-Theory, the Simons
Center 2012 Program on String Compactification, and the String Vacuum
Project 2012 meeting (Simons Center).}
}
\author{Gordon Kane}
\author{Ran Lu}
\author{Bob Zheng}

\affiliation{Michigan Center for Theoretical Physics, University of Michigan, Ann Arbor, MI 48109 USA}

\date{\today}

\begin{abstract}
The August 2011 Higgs mass prediction was based on an ongoing six year
project studying M-theory compactified on a manifold of $G_{2}$ holonomy,
with significant contributions from Jing Shao, Eric Kuflik, and others,
and particularly co-led by Bobby Acharya and Piyush Kumar. The M-theory results include:
stabilization of all moduli in a de Sitter vacuum; gauge coupling unification;
derivation of TeV scale physics (solving the hierarchy problem); the derivation that
generically scalar masses are equal to the gravitino mass which is larger
than about 30 TeV; derivation of the Higgs mechanism via radiative
electroweak symmetry breaking; absence of the flavor and CP problems, and the accommodation of string axions. $\tan\beta $ and the $\mu $ parameter are part of the theory
and are approximately calculated; as a result, the little hierarchy problem is greatly
reduced. This paper summarizes the results relevant to the Higgs mass
prediction. A recent review \cite{1204.2795} describes the program more broadly. Some of
the results such as the scalar masses being equal to the gravitino mass and
larger than about 30 TeV, derived early in the program, hold generically for
compactified string theories as well as for compactified M-theory, while some other results may or may
not. If the world is described by M-theory compactified on a $G_{2}$
manifold and has a Higgs mechanism (so it could be our world) then $M_{h}$
was predicted to be $126\pm 2$ GeV before the measurement. The
derivation has some assumptions not related to the Higgs mass, but involves no free
parameters.
\end{abstract}
\maketitle

\section{INTRODUCTION}

How should one proceed to make testable predictions from M-theory or string theory? \
Obviously it is first necessary to compactify the theory to have six or
seven small curled up dimensions. Perhaps someday the theory will tell us
which corners of string theory to compactify, along with the matter content and gauge group upon compactification,but until then it is good to pick out likely-looking
manifolds and try them. Although there may be many string vacuum solutions, finding examples which resemble our world at present levels of knowledge is possible. The compactified theory will generically have
moduli that describe the shapes and sizes of the curled up dimensions, and
it is necessary to calculate the potential for the moduli and to stabilize
them. 

We can divide all compactified string/M-theories into two classes:

\begin{itemize}

\item Some have de Sitter vacua, TeV scale physics, a Higgs mechanism via
radiative electroweak symmetry breaking (EWSB), no cosmological or
phenomenological contradictions with data, etc. If our world is described
by a compactified string/M theory it will look like these, and we can then
calculate the Higgs boson mass and properties (and predictions for LHC physics, rare
decays, dark matter, etc.). It turns out it is not hard to find such a
theory. In the following we describe the details.

\item The rest.

\end{itemize}

We describe the calculation of the Higgs boson mass $M_{h}$ for the case of M-theory compactified
on a manifold of G$_{2}$ holonomy in a fluxless sector, since for that case many of the
calculations have been explicitly done. Some of the results depend only on
a few generic features and may hold more generally for other corners of
string theory. For the compactifications with the Higgs mechanism we
compute the ratio $M_{h}/M_{Z}$ precisely. Given the experimental value of $M_Z$, we find $M_{h}$ = 126 $\pm $ 2 GeV. The results were first presented at the International String Phenomenology Conference in August 2011 \cite{link}. The
resulting theory has a decoupling Higgs sector, so the Higgs boson
with the smallest eigenvalue of the CP-even mass matrix is Standard Model-like.

\section{BRIEF HISTORY OF M-THEORY COMPACTIFICATION}

Since the history of M-theory compactifications is poorly known, we briefly
summarize it. Soon after M-theory was developed in 1995, Papadopooulos and
Townsend \cite{th/9506150} showed that compactifying the effective 11-D supergravity theory on a manifold with $G_2$ holonomy preserved N=1 supersymmetry. Then a series of papers
developed the structure of the compactified theory:

\begin{itemize}
\item Papadopoulos and Townsend: realized that 4D compactification with $G_2$ holonomy
preserves N=1 supersymmetry \cite{th/9506150}.

\item Acharya: demonstrated how non-abelian gauge fields are localized on 3 cycles near ADE singularities \cite{th/9812205}, \cite{th/0011089}.

\item Atiyah and Witten: studied M-theory compactifications on $G_2$ manifolds near conical singularities \cite{th/0107177}.

\item Atiyah, Maldacena and Vafa: established a duality between type II-A strings and M-theory \cite{th/0011256}.

\item Acharya and Witten: showed that chiral fermions can be supported at points
with conical singularities \cite{th/0109152}.

\item Witten: discussed possible ways to embed the supersymmetric Standard Model in $G_2$ compactifications while solving the triplet-doublet splitting problem \cite{ph/0201018}.

\item Beasley and Witten: computed the effective 4-D Kahler potential upon Kaluza-Klein reduction of M-theory on $G_2$ \cite{th/0203061}.

\item Friedmann and Witten: studied threshold corrections to gauge coupling unification in $G_2$ compactifications \cite{th/0211269}. 

\item Lukas and Morris: computation of the gauge kinetic function resulting from $G_2$ compactifications \cite{th/0305078}.

\item Acharya and Gukov: a review with a good summary of known
results about singularities, holonomy and supersymmetry in $G_2$ compactifications \cite{th/0409191}.

\end{itemize}

Based on these developments, Acharya and Kane started the program to study in the detail the effective four dimensional theory which would arise from such a compactification in 2005 with several people, including early major
contributions from Konstantin Bobkov and Piyush Kumar, and later ones from
Eric Kuflik, Piyush Kumar, Jing Shao, Ran Lu, Scott Watson, and Bob Zheng. The resulting
papers covered a range of topics:

\begin{itemize}

\item \emph{M-Theory Solution to Hierarchy Problem} \cite{th/0606262}.

\item \emph{Explaining the Electroweak Scale and Stabilizing Moduli in M Theory.} Predicted the gravitino and squark masses, argued the gaugino masses
were suppressed, and showed gravity mediation was generic \cite{th/0701034}.

\item \emph{The G(2)-MSSM: An M Theory Motivated Model of Particle Physics}. Calculated
the spectrum resulting from $G_2$ compactifications in detail \cite{0801.0478}.

\item \emph{Non-thermal Dark Matter and the Moduli Problem in String Frameworks}. Studied the moduli, showed the resulting cosmological history was non-thermal with
nonthermal Dark Matter, and showed moduli masses were larger than about 30 TeV \cite{0804.0863}. 

\item \emph{CP-violating Phases in M-theory and Implications for EDMs}. Showed CPV Phases in M-theory could be rotated away so weak CPV constraints are satisfied, and calculated EDMs \cite{0905.2986}.

\item \emph{String Theories with Moduli Stabilization Imply Non-Thermal Cosmological History, and Particular Dark Matter}. Related moduli and gravitino masses and showed that the gravitino is generically heavier than the lightest moduli, and thus squark masses are of order the gravitino mass and larger than about 30 TeV \cite{1006.3272}. 

\item \emph{An M-Theory Solution to the Strong CP Problem and Constraints on the Axiverse}\cite{1004.5138}.

\item \emph{Identifying Multi-Top Events from Gluino Decay at the LHC} \cite{0901.3367}.

\item \emph{Theory and Phenomenology of $\mu$ in M-theory} \cite{1102.0556}.

\item \emph{String-motivated approach to the Little Hierarchy Problem} \cite{1105.3765}.

\item \emph{Higgs Mass Prediction in Realistic String/M-theory Vacua} \cite{1112.1059}.

\item \emph{Compactified String Theories - Generic Predictions for Particle Physics}. A comprehensive review of recent progress made in string/M-theory compactifications, with a focus on phenomenological consequences \cite{1204.2795}.

\end{itemize}

In general we have emphasized compactification and stabilizing moduli and
supersymmetry breaking, focusing on predictions for LHC and dark matter
physics, rather than the explicit embedding of quark and lepton matter in the compactified theory.

\section{ASSUMPTIONS}

In the following we make certain assumptions, all standard and expected to hold. None are
directly related to the Higgs sector. Some can be derived or checked
separately already or later. At the present time each compactification has
to be treated separately, and we focus on a particular one: M-theory
compactified on a manifold of $G_2$ holonomy, with minimal supersymmetric Standard Model (MSSM) matter and gauge content. It will be clear from the derivation that the
results are likely to hold more widely. Here we list the assumptions:

\begin{itemize}

\item Assume our world is described by M-theory compactified on a manifold of $G_2 $ holonomy in the fluxless sector.

\item Assume that the Hubble parameter H at the end of inflation was larger than the gravitino mass $m_{3/2}.$

\item Assume the effective supergravity approximation is valid after compactification (standard in studying the 4-D limit of string theories).

\item Assume the cosmological constant (CC) problem is decoupled from other physics (see below).

\item Assume there is a top quark with yukawa coupling $\approx $ 1.

\item Assume that the SM embeds into a GUT group. This naturally leads to gauge coupling unification at the GUT scale, as is well known.

\item Assume that the expected form of the Kahler potential and  gauge kinetic function derived in previous works \cite{th/0203061,th/0305078} are approximately correct.

\item Assume that a very large value of $\mu$ is avoided by using Witten's mechanism \cite{ph/0201018} (see below).

\item Assume that the gauge group and matter content below the compactification scale is that of the MSSM.

\end{itemize}

The framework thus defined makes a number of predictions for observables that will be tested in the near future. As discussed in references \cite{th/0701034} and \cite{0801.0478}, the resulting theory has stabilized moduli and no free parameters, with supersymmetry breaking via hidden sector gaugino condensation and hidden sector chiral fermion condensation. The
soft-breaking Lagrangian and the $\mu $ parameter are then calculated in the
compactified theory; the gravitino mass is calculated to be $\mathcal{O}$(50 TeV)\cite{0801.0478,0810.3285}. The first and last assumptions can be replaced with alternatives and
the calculation repeated. The next to last is natural or even required in
the M-theory context, but could be different in other corners of string
theory. The rest of the assumptions are very likely to be true. We comment
below on the status of the CC assumption and the $\mu $ embedding. 

The state of M/string theory is such that predictions from compactified
M/string theories can be made and tested for particular compactifications,
and in some cases for large classes of compactifications. Here we focus
on the prediction of the Higgs boson mass, and we begin by assuming the
matter and gauge group content at the compactification scale is that of the
MSSM. Later we remark briefly on alternatives. Since the MSSM gauge and matter content gave the
correct prediction, the question of alternatives is subtle.

Witten suggested in 2002 that the $\mu $ parameter in M-theory be set to
zero in the superpotential by a discrete symmetry \cite{ph/0201018}. We recently recognized that
moduli stabilization would break this discrete symmetry \cite{1102.0556}. Consequently the
value of $\mu $ which would naively be of order $m_{3/2}$ is in addition
suppressed by a ratio of a typical moduli vacuum expectation value (vev) to the Planck scale, a number
of order 0.05-0.1. In the following we assume $\mu \lesssim 0.1 m_{3/2}$.

The usual conditions for electroweak symmetry breaking (EWSB) obtained by minimizing the Higgs potential are:

\begin{align}
&\frac{M_{Z}^{2}}{2}=-\mu^{2}+\frac{M_{H_{d}}^{2}-M_{H_{u}}^{2}\tan ^{2}\beta }{\tan ^{2}\beta -1}\notag\\
&2B\mu =\sin 2\beta(M_{H_{u}}^{2}+M_{H_{d}}^{2}+2\mu ^{2}).
\end{align}

\begin{figure}[t!]
    \begin{center}
        \includegraphics[height=10cm, width=12cm]{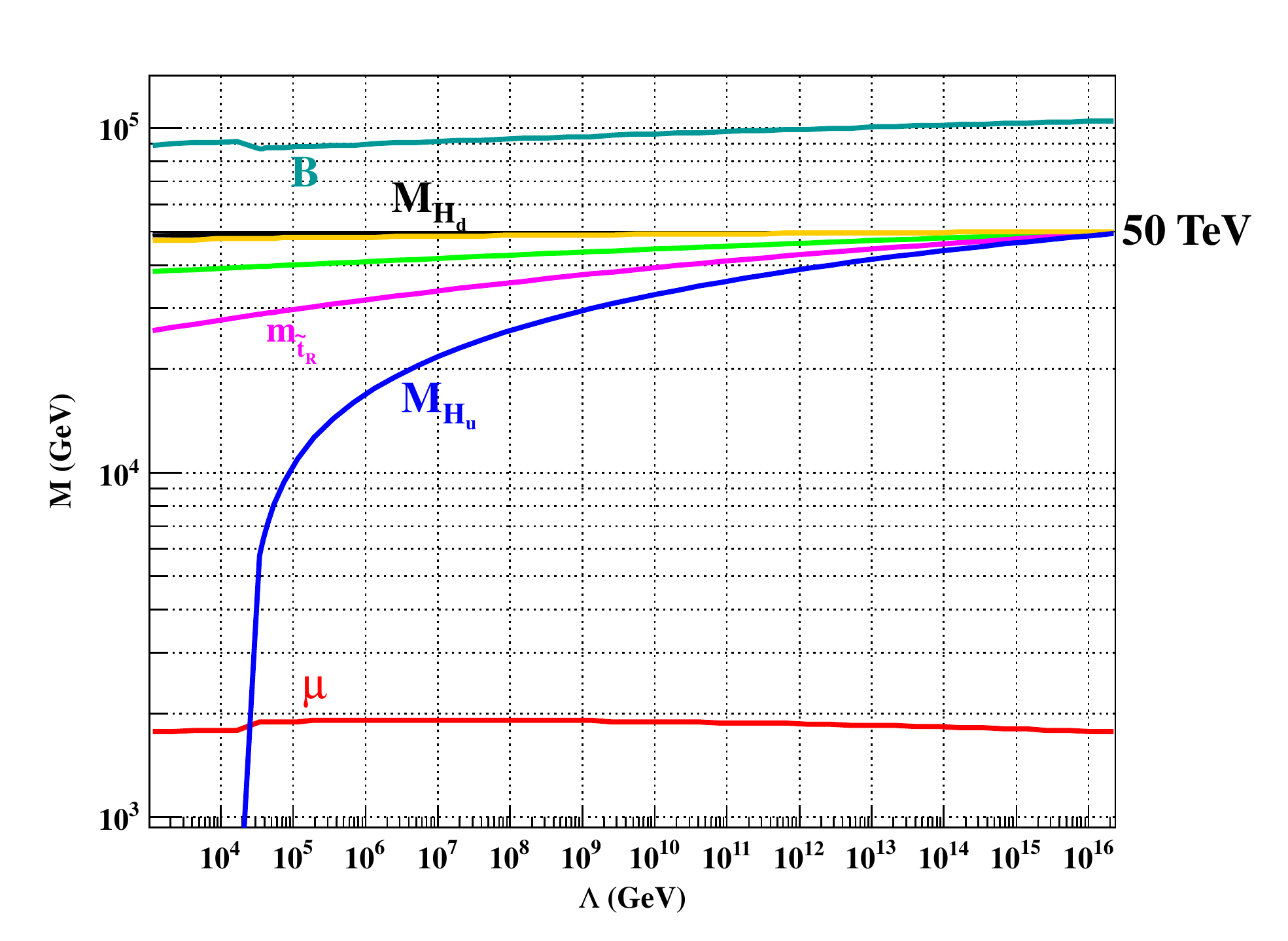}
    \end{center}
    \caption{Renormalization group evolution of $\mu$, $B$ and scalar soft masses.}
    \label{fig:rge}
\end{figure}

$M_{H_{u,d}}$ are soft breaking scalars whose values are set by the gravitino mass, and are therefore $\gtrsim$ 30 
TeV at the unification scale. In renormalizing the soft breaking parameters down to the TeV scale, the large top yukawa causes $M_{H_{u}}$ to run to small values, of order a TeV, while $M_{H_{d}}$ doesn't run much due to the small bottom yukawa. $B$ runs down about 15\% as shown in Figure \ref{fig:rge}. When the superpotential does not contribute to $\mu$ or $B\mu$, and the moduli $F$-terms are supressed compared to the hidden sector meson condensate $F$-terms, and the visible sector Kahler metric and the Kahler potential $H_u H_d$ coefficient is independent of the meson field, then this implies $B\approx 2M_{3/2}$ \cite{0801.0478,1102.0556}. For $\tan \beta $ not too small, $\sin 2\beta \approx
2/\tan \beta .$ Then $\tan \beta \approx M_{H_{d}}^{2}/B\mu \approx
m_{3/2}^{2}/B\mu $ so finally we expect $\mu \tan \beta \approx m_{3/2}/2$
as a constraint on M-theory vacua with proper EWSB. Including running effects on $B$
moves the $2~$in the denominator closer to about $1.7.$ These RGE effects are discussed in more detail in \cite{1105.3765}; they are obtained assuming universal soft breaking masses $M_0 = m_{3/2}$ and large universal soft breaking trilinears $A_0 \gtrsim m_{3/2}$ at the unification scale. Such conditions on the soft breaking parameters are automatically satisfied for the M-theory compactifications which we consider \cite{0801.0478}. The M-theory compactification generically predicts a gravitino mass $m_{3/2} \sim 50$ TeV \cite{0801.0478,0810.3285}. The Higgs mass shifts by about $\pm 1.5$ GeV for a change in the gravitino mass of a factor of two in either direction.

We assume no solution to the cosmological constant problems in a
particular vacuum, and that they are solved by some other physics, which we
assume is basically decoupled or orthogonal to the compactified M/string
theory physics we are considering. Not solving the cosmological constant
problem seems unlikely to prevent us from calculating the Higgs boson
mass, and solving the cosmological constant problems seems unlikely to
help us calculate the Higgs boson mass. Of course we cannot be sure of
this until the cosmological constant problems are solved, but we proceed
under these assumptions.

There is a very general relation between the lightest eigenvalue of the
moduli mass matrix and the gravitino mass, first derived in refs \cite{th/0411183,th/0602246} and
later rederived in \cite{1006.3272}. As long as supersymmetry breaking is involved in
stabilizing at least one modulus, then the moduli mix with the scalar goldstino,
whose mass is generically the gravitino mass. Consider the moduli mass matrix (whose explicit form need not be known for the following). Then the smallest eigenvalue of the
full matrix is smaller than any eigenvalue of the diagonal submatrices, and
one can show:

\begin{equation}
M^2_{\mathrm{mod}, \min} \lesssim {m_{3/2}^2} \left(2 + \frac{\left|r\right|}{M_{pl}^2}\right)
\end{equation}

where $r$ is the sectional curvature in the space of scalar fields. If no new scales are introduced then
normally $\frac{\left\vert r\right\vert}{M_{pl}^2} \lesssim 1.$

In the early universe as the Hubble scale $H$ decreases, moduli begin to
oscillate in their potential, and quickly dominate the energy density of the
universe. When $H$ is of order the moduli decay width, $\Gamma _{\func{mod}%
}\approx M_{\func{mod}}^{3}/M_{Pl}^{2}$, the moduli begin to decay. This must
occur before nucleosynthesis in order not to disrupt the successful predictions of big bang nucleosynthesis (BBN), which imply $M_{%
\func{mod}}\gtrsim 30$ TeV \cite{ph/9906527,0804.0863}. By the above result this implies $m_{3/2}\gtrsim
30$ TeV. In supergravity generically squark and slepton masses will be nearly equal to
the gravitino mass, so squarks are heavier than about $30$ TeV \cite{1006.3272}; this makes them too heavy to be observed at the LHC. This \
prediction is very general for compactified M/string theories. 

Sometimes people discuss escaping this conclusion by having a carefully
timed late inflation. Many authors have tried to find a way to arrange
this, and failed. The recent paper of Fan, Reece, and Wang is a thorough
study \cite{1106.6044}. There is certainly no known generic way to establish such a cosmological scenario, and if there is
a way it will be highly fine tuned.

\section{THE HIGGS SECTOR}

A Standard Model Higgs boson is severely fine-tuned because of the
quadratically divergent radiative corrections to its mass. In
the MSSM there are two Higgs doublets, both for anomaly cancellation
and to ensure both up and down type quarks and leptons get mass. When we refer to the Higgs boson mass $M_h$, we mean the mass of the smallest eigenvalue of the tree level CP
even mass matrix. The Higgs potential is calculable, leading to a mass
matrix:

\begin{equation}
\left(
\begin{array}{cc}
M_{H_{u}}^{2}+\mu ^{2} & -B\mu \\
-B\mu & M_{H_{d}}^{2}+\mu ^{2}%
\end{array}
\right)
\end{equation}

\begin{figure}[t!]
    \begin{center}
        \includegraphics[width=12cm,height=8cm]{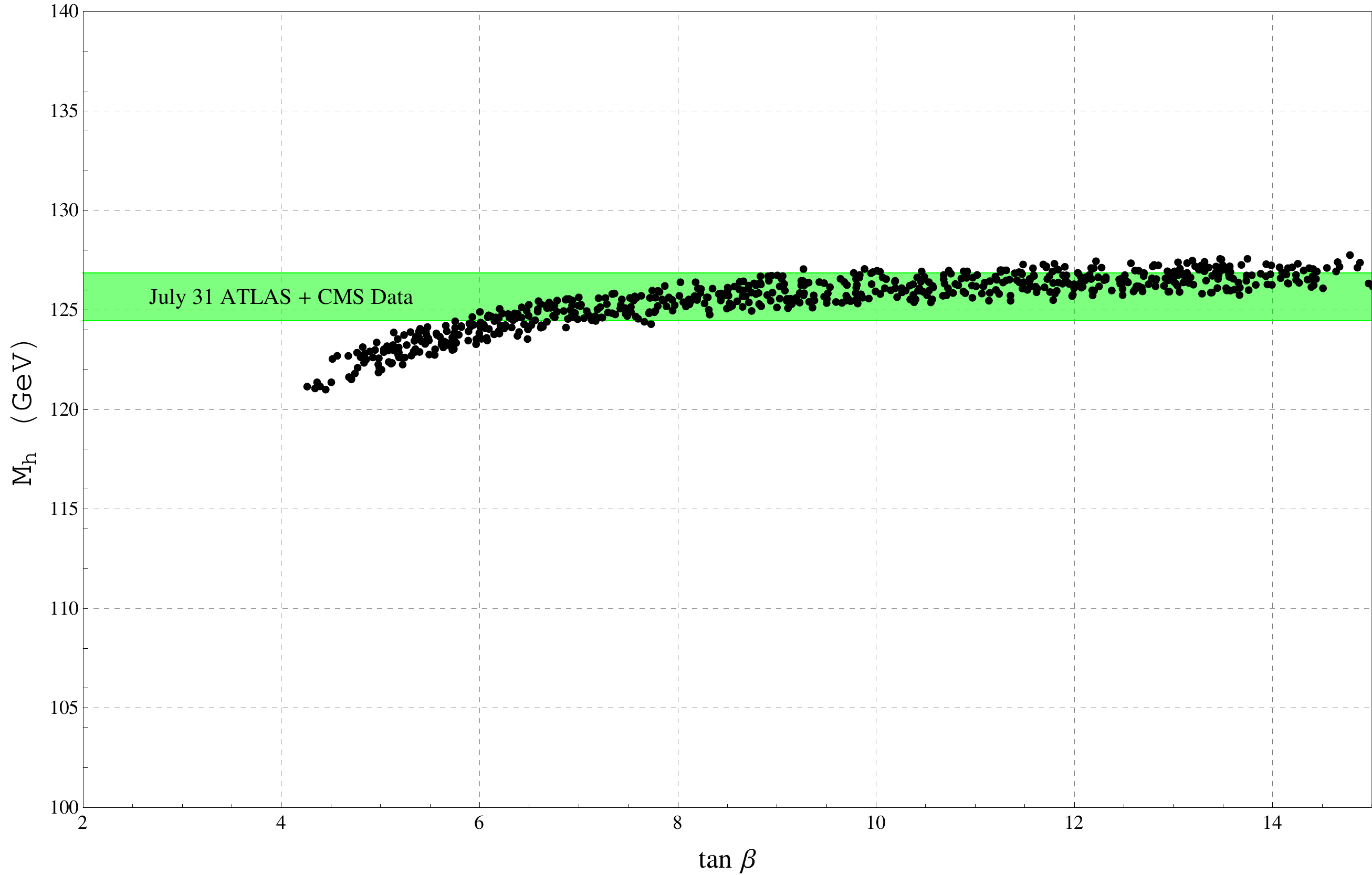}
    \end{center}
    \caption{Compactified M/string-theory prediction of the Higgs boson mass for $\mu \lesssim 0.1 m_{3/2}$ and $m_{3/2} = 50$ TeV, given the stated assumptions. For such values of $\mu$, requiring EWSB gives $6 \lesssim \tan\beta \lesssim 15$. The dots show all solutions with EWSB.}
    \label{fig:higgs1}
\end{figure}

At the unification scale there is no EWSB. As $M_{H_{u}}^{2}$ runs down (as discussed in the preceding section),
a negative eigenvalue emerges and two non-zero Higgs vev's are generated, $v_u$ and $v_{d}$. Their ratio $\tan \beta = v_{u}/v_{d}$ does not
exist at the unification scale, and becomes meaningful only when the vev's are
generated. At the unification scale all of the MSSM parameters are fixed by the
theory. Then one carries out the RGE running down to the TeV scale, finding all solutions
with EWSB. The running of several of the soft-breaking terms is shown in
Figure \ref{fig:rge}.We use both available software such as SOFTSUSY \cite{ph/0104145}, and our own 
``match-and-run" method inspired by the work of \cite{1108.6077}, matching at the renormalization scale equal to the geometric mean of the stop masses $Q=\sqrt{M_{\tilde{t_1}}M_{\tilde{t_2}}}$
with two-loop RGE's. SOFTSUSY should work well since the
scalar masses are not larger than tens of TeV, and we check that the two
methods agree. The main sources of imprecision are the experimental uncertainty of the top quark mass,
which gives a $\pm 0.8$ GeV uncertainty in the Higgs mass,
and the experimental uncertainty in $\alpha _{s}$, which gives a $\pm 0.3$ GeV uncertainty in the Higgs mass \cite{1112.1059}. The compactified M/string theory gives trilinears that
are about 1.4 times $m_{3/2}$ at the unification scale, and we check that a variation in the
trilinears of order 25\% has a small effect on the Higgs mass. Similarly,
the gluino should have a mass of about a TeV, and we check that a 20\% variation in the
gluino mass does not change the Higgs mass by more than 0.3 GeV. The points in Figure \ref{fig:higgs1}
show all solutions with EWSB allowing for variations of the top yukawas, $%
\alpha _{3}$, trilinears, gluinos. As shown above, $\tan \beta $ is
expected to be about $\tan \beta \approx m_{3/2}/1.7\mu ,$ so if $m_{3/2}\approx 50$ TeV and $%
\mu \approx 3$ TeV, then $\tan \beta \approx 10,$ resulting in $M_{h}\approx 126\pm 2$ GeV if we allow for some variation in $m_{3/2}$.

\begin{figure}[h!]
    \begin{center}
        \includegraphics[width=12cm,height=8cm]{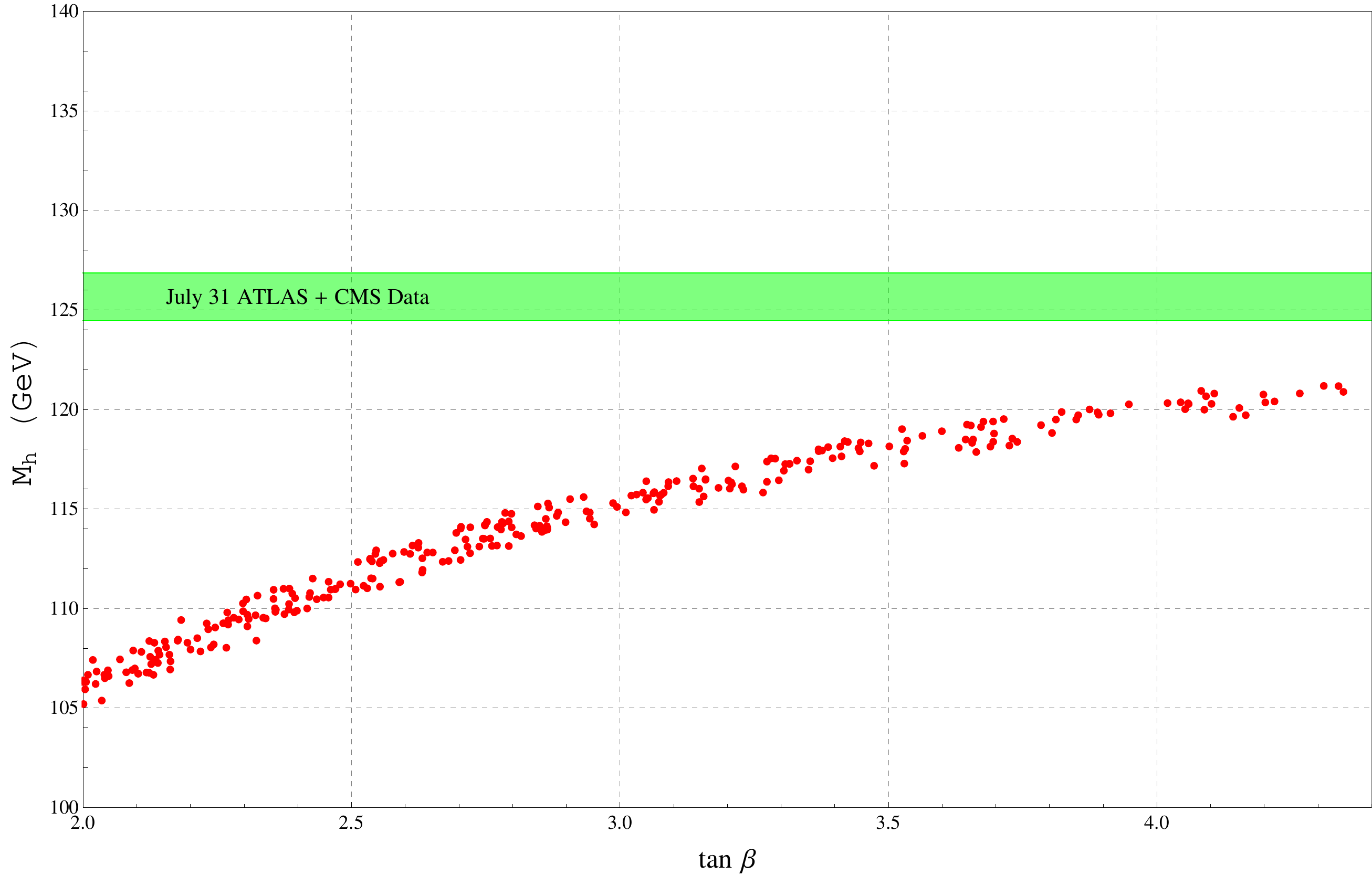}
    \end{center}
    \caption{Compactified M/string-theory prediction of the Higgs boson mass for $\mu \sim m_{3/2}$ and $m_{3/2}= 50$ TeV, given the stated assumptions (modulo the assumption regarding $\mu$). For such values of $\mu$, requiring EWSB gives $\tan\beta \lesssim 4$.}
    \label{fig:higgs2}
\end{figure}

Figure \ref{fig:higgs2} shows solutions with $\mu \approx m_{3/2}.$ Note that with respect to the experimental result, there is a
significant preference for the solutions with $\mu $ suppressed by an order
of magnitude or more below $m_{3/2}$ by the broken discrete symmetry.

\section{MITIGATING THE LITTLE HIERARCHY PROBLEM}

As previously discussed, the soft -breaking Lagrangian predicted by the compactified theory always
has trilinears $A_{0}\gtrsim M_{0}$, the degenerate tree level scalar mass.
Then $M_{H_{u}}$ runs down to about a TeV, and $M_{H_{d}}/\tan
\beta $ and $\mu $ can be similarly in the TeV region. A helpful way to
see this is to express the low scale $M_{H_{u}}^{2}$ in terms of the high
scale soft terms, which gives \cite{1105.3765}:

\begin{equation}
M_{H_{u}}^{2}\approx f_{M}M_{0}^{2}-f_{A}A_{0}^{2}.
\end{equation}

The $f_{i}$ have values of order $0.1$, so with the unification scale scalars
and trilinears similar in size there is a double suppression for $%
M_{H_{u}}^{2}.$Thus even though the natural scale of the scalars at the unification scale is of
order $50$ TeV, the natural vev of the Higgs is of order $1-2$ TeV. There
is still a residual little hierarchy problem, but it is not worse, and
perhaps a little better, for the compactified M/string theory compared to the typical tuning required for $M_h \sim 126$ GeV in the MSSM \cite{1112.2703}. Since no well motivated approach can calculate the observed Higgs vev without large tuning, the compactified M/string approach where the results are derived and not tuned should be considered to
have significant success here. This result depends on having large trilinears $A_0 \gtrsim m_{3/2}$ as required by the compactification, and thus will not hold e.g. if there is an approximate R-symmetry which suppresses the trilinears.

In this regard, it is interesting that some time ago it was pointed out \cite{0903.1024}
out that physics did not change if the Higgs vev were several times larger.

\section{THE DECOUPLING LIMIT}

In the decoupling limit where the scalars are heavy, the other MSSM Higgs bosons
H,A, H$^{\pm }$ have masses of order $m_{3/2}$. More importantly, the lightest Higgs decay
branching ratios are quite similar to the Standard Model Higgs, though of
course it could not be a purely Standard Model Higgs since we have a solution to the hierarchy problem. While the calculated mass and the
branching ratios are logically not connected, constraints on any mechanisms
that could change the Higgs decay branching ratios are quite tight in the compactified M-theory framework.

For instance, there has been much recent interest in increasing the $h \rightarrow \gamma \gamma$ branching ratio of the SM-like Higgs \cite{1112.3336,1205.5842,1206.1082,1207.2473,1207.4235,1207.7062}, motivated by recent hints in the LHC data. The most straightforward way of obtaining an enhanced $h \rightarrow \gamma \gamma$ signal is via additional BSM matter running in the $h \rightarrow \gamma \gamma$ loop; these BSM particles must be light enough to have contributions comparable to the SM particle loops, and should be charged and uncolored to avoid disturbing the $g g \rightarrow h$ production mode which would affect the other branching ratios. In the unconstrained MSSM, it is possible for light sleptons \cite{1112.3336} (in particular a light stau \cite{1205.5842}) to provide such an enhancement; however this will not work in compactified string/M-theories where sleptons have masses $\gtrsim 30$ TeV. 

Another way of potentially increasing $h \rightarrow \gamma \gamma$ which is consistent with heavy scalar superpartners involves additional vector-like leptons in the theory \cite{1206.1082,1207.4235,1207.7062}, which can give up to a 100$\%$ enhancement for sufficiently large couplings to the Higgs. However, sufficiently large couplings will inveitably induce a large shift in the Higgs mass \cite{0910.2732,1207.4482}, which will ruin the succesful prediction of the Higgs mass in compactified M-theories. It is possible that the effect of the additional couplings in the RGE's for $M_{H_u}$ and $M_{H_d}$ will allow for a smaller $\tan \beta$ to compensate for the positive shift in $M_h$ induced by these couplings. However, we find that even in this scenario the $h \rightarrow \gamma \gamma$ enhancement can at most reach $\approx 15\%$, as larger couplings of the vector-like leptons would drive the vector-like slepton masses tachyonic upon running down to the TeV scale.  

\section{EXTENDED GAUGE GROUP AND MATTER AT COMPACTIFICATION}

We comment briefly on variations in the gauge group and matter content upon compactification that could potentially effect the Higgs mass calculation. One class of gauge extensions involves an additional gauge group $\mathcal{H}$ under which $H_u$ and $H_d$ are charged, which is Higgsed at some scale in an $F$-flat and $D$-flat direction so that breaking $\mathcal{H}$ does not break supersymmetry. This gives a contribution to the Higgs mass which is parametrically $\delta M_{h}^2 \sim g_{X}^2 v^2 \left(1 + \Lambda_{H}^2/m_S^2\right)^{-1}$\cite{ph/0309149,ph/0409127,1207.0234}. Here $v$ is the SM Higgs vev, $g_X$ is the gauge coupling for $\mathcal{H}$, $\Lambda_H$ is associated with the symmetry breaking scale of $\mathcal{H}$ and $m_S$ is the soft mass of the scalar field whose vev is responsible for the breaking. 

Thus if the breaking scale of $\mathcal{H}$ is not too far from the scale of the soft-masses of the theory, there may be a sizable non-decoupling shift to $M_h$. However, UV completions of such models require the generation of a new scale $\Lambda_H$ which is a priori unrelated to the supersymmetry breaking scale. Furthermore, engineering superpotentials with $F$-flat directions typically involve introducing multiple exotic superfields, making their embedding into simple GUT groups difficult and non-minimal. There is no known way to naturally motivate such models via compactified string theories such that $\Lambda_H \sim \mathcal{O}(m_{3/2})$, so we will not consider them further\footnote{In certain classes of compactified string theories, $\Lambda_H$ can be generated via the Stuckelberg mechanism for a $U(1)_X$. However, this generically results in $\Lambda_H \sim M_{pl}$, so the shift to $M_h$ will be negligible. A notable exception is given in \cite{ph/0205083}.}.

Another class of gauge extensions involves an extended gauge symmetry which is Higgsed in non $F$-flat and $D$-flat directions. This symmetry breaking can occur for example by the soft mass squared of some SM singlet superfield running negative via RGE effects; in these cases, the breaking scale of the gauge extension is directly related to the scale of the soft scalar masses \cite{ph/9511378,ph/9703317}. Such gauge extensions can be embedded in straightforward ways into simple GUT groups such as $SO(10)$ or $E_6$. 

If there are no superpotential couplings between $H_u/H_d$ and exotic matter, and the Higgs doublets are charged under the gauge extension, the contribution to $\delta M_{h}^2$ comes purely from the $D$-term and is parametrically $\delta M_h^2 \sim - g_{X}^2 v^2 \left(v^2/s^2\right)$, where $s \sim \mathcal{O}(m_{3/2})$ is the vev of the field which breaks the extended gauge symmetry. In the compactified string theory context with $m_{3/2} \gtrsim 30$ TeV, $v/s \ll 1$ so $\delta M_h^2$ would be negligible in these models. Therefore such gauge extensions could potentially be consistent with the compactified M-theory calculation of a $126$ GeV Higgs. However, if the field which Higgses the extended gauge symmetry has no large superpotential couplings, the tachyonic running of the soft mass squared must be driven by the supertrace of the extended gauge group, which requires a significantly non-universal soft mass spectrum. This is discussed in \cite{1005.5392} for the case of $SO(10)$, where the right-handed sneutrino vev breaks the extended gauge symmetry. Such a soft mass hierarchy is difficult to realize in the case of supergravity mediated supersymmetry breaking which generically gives universal scalar masses.

If the SM singlet field which Higgses the extended gauge symmetry does have superpotential couplings to the Higgs doublets, $W \supset \lambda S H_u H_d$, then there can be a non-decoupling shift $M_h^2$ which survives even in the $v/s \sim v/m_{3/2} \ll 1$ limit \cite{ph/9703317}. This effect has been closely examined in the context of $E_6$ \cite{Haber:1986gz,Drees:1987tp,ph/9804428,1206.5028}. However, such a gauge symmetry would also forbid the bare Kahler $\mu$ term, requiring us to abandon Witten's solution to the $\mu$ problem developed in \cite{ph/0201018,1102.0556} which naturally gives $\mu \sim \mathcal{O}\left(0.1 m_{3/2}\right)$. Instead, we would have an NMSSM-like solution to the $\mu$ problem, $\mu = \lambda s$; in order to reproduce $\mu \sim \mathcal{O}\left(0.1 m_{3/2}\right)$, $\lambda$ must be tuned to be small. Furthermore, both of the preceding classes of gauge extensions would induce shifts in $M_{H_u}^2/M_{H_d}^2$ proportional to $g_{X}^2 s^2$ upon symmetry breaking, which would contribute to the little hierarchy problem. Thus there does not seem to be a straightforward way to extend the gauge symmetry of the MSSM in a way that does not significantly change the $M_h \sim 126$ GeV result computed in \cite{1112.1059} while maintaining natural EWSB as described in \cite{1105.3765}.

\section{SUPPRESED GAUGINO MASSES AND SOME LHC PREDICTIONS}
 
\begin{figure}[t!]
    \begin{center}
        \includegraphics[width=12cm]{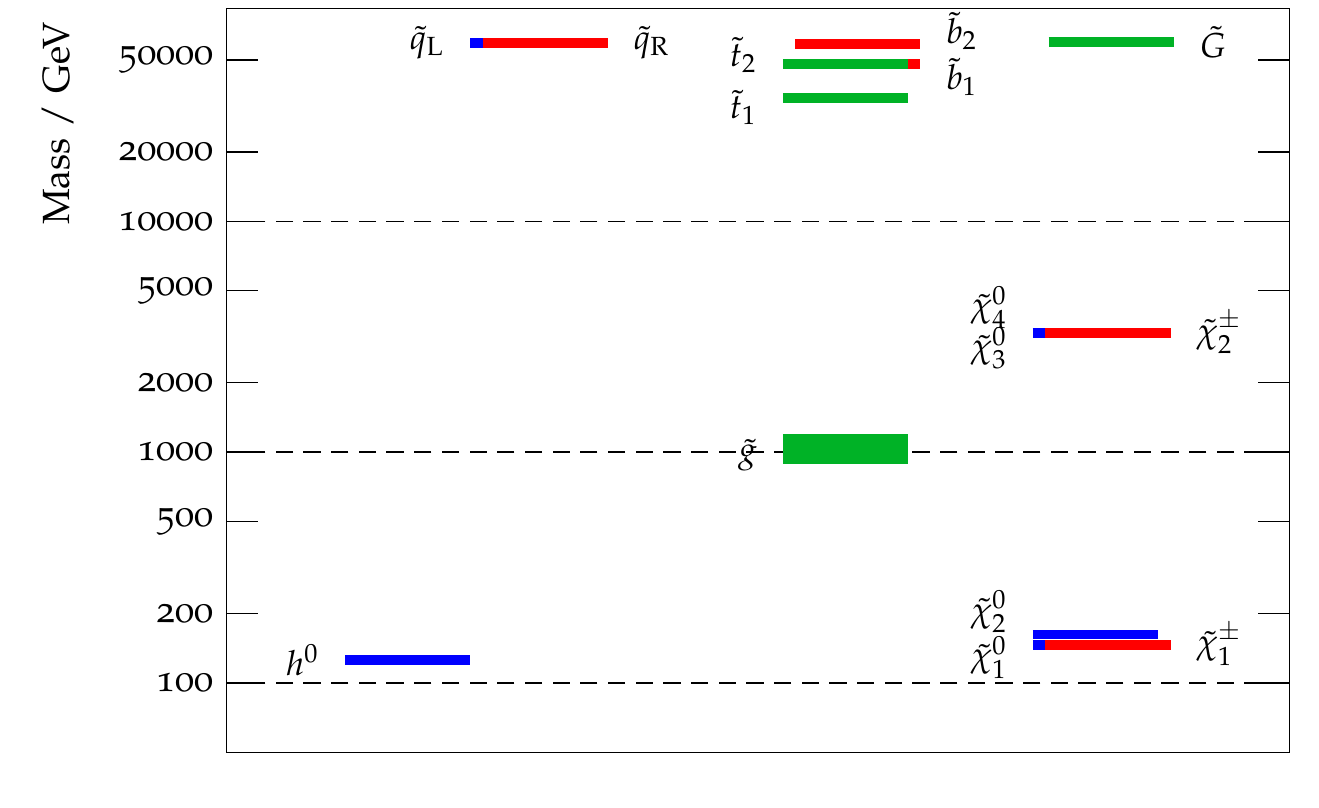}
    \end{center}
    \caption{Typical spectrum of $G_2$-MSSM, with $m_{3/2} = 50$ TeV.}
    \label{fig:spectrum}
\end{figure}

We have explained that one very robust prediction if the world is described
by a compactified M/string theory is that scalars are too heavy to be
observed at the LHC. On the other hand, gluino, chargino, and neutralino masses are suppressed by a
dynamical mechanism. Note that this suppression cannot be due to an
R-symmetry because the trilinear couplings are of order the gravitino mass.
The suppression occurs because the dominant supersymmetry breaking is due to
the $F$-term from the chiral fermion condensation to a meson, with $%
F_{meson}\sim m_{3/2}M_{pl},$ while the gaugino masses are given by

\begin{equation}
M_{1/2}^{a}=\frac{%
\sum_{i}F^{i}\partial _{i}f^a_{vis}}{2i\func{Im}(f^a_{vis})}
\end{equation}

where $F_{i}$ are the $F$ terms for the moduli and $f_{vis}$ is the visible
sector gauge kinetic function \cite{th/0305078}, and the index $a$ runs over the bino, winos and gluinos. Thus the mass scale of the gauginos is set by
the suppressed moduli $F$ terms rather than the dominant meson $F$ terms. \
The moduli $F$ terms are generically suppressed by the volume of the
manifold, which is of order $\sim \alpha _{GUT}$ in Planck units. Therefore a typical spectrum would
look like the spectrum shown in Figure \ref{fig:spectrum} \cite{0801.0478}.

\section{ENHANCED GLUINO DECAYS TO THIRD FAMILY}

Since the gauginos generically have suppressed masses while other scalars are beyond the reach of the LHC, the phenomenology of the gluino is the most interesting topic to study at the LHC. In general, the gluino decays to all the SM quarks through an off-shell squark state. Therefore the branching fraction to a given quark is inversely proportional to the mass of its squark superpartner to the 4th power $(M_{\tilde{q}})^{-4}$, so the decay channels through relatively light squarks are significantly enhanced. From Figure \ref{fig:spectrum}, we can see that the right handed stop is the lightest, followed by the left handed stop and sbottom pair. Numerically this would enhance the decays to the third family quarks by a factor of 2-10, and the total branching fraction to the third family can be as large as 50\%. Thus final states with a lepton and multiple b-tagged jets will be powerful probes of gluino pair production at the LHC \cite{0901.3367,1101.1963}.

\section{CONCLUDING REMARKS}

The correct  prediction of the Higgs boson mass described here is not an
accident. Any compactified M-theory or string theory with moduli, in which some or all moduli are stabilized by mechanisms that break supersymmetry such as
gaugino condensation, will generically havea supergravity field theory as its low energy limit, with gravitinos and
therefore scalars in the multi-TeV region to be consistent with cosmological constraints. This results in a supersymmetric
decoupling Higgs sector with one light Higgs boson that is like the Standard
Model Higgs boson -- though of course not the pure SM Higgs boson since that has
a quadratically divergent mass and any associated physical observables are not defined.  A
complete theory requires that the $\mu $ parameter and $\tan \beta $ be
included in the theory. In compactified M-theory $\mu $ should be
suppressed by moduli stabilization to somewhat over an order of magnitude
below $m_{3/2}$ and $\tan \beta $ should be in the $10-15$ range, in which
case the Higgs mass is $125$ GeV to an accuracy of about $1.5\%.$ The M-theory
calculation has no free parameters. It has some assumptions not related to
the Higgs sector, and M/string theory predictions still have to be done
separately for each compactification. In other corners of string theory
there have been arguments for $\mu \sim m_{3/2}$ \cite{1009.0905,1012.4574,1102.3595}, which we would expect to
lower the Higgs mass by several GeV due to a lower $\tan\beta$, but this situation has not yet been fully
resolved.

Since the theory predicts heavy squarks and sleptons, they should not be
found at the LHC. The dynamics of supersymmetry breaking implies that gluinos
should be in the $\mathcal{O}(1)$ TeV region, with one chargino and two neutralinos
lighter than a TeV. The gluino and these lighter gauginos are detectable at the LHC.The
heavy moduli will decay before nucleosynthesis, introducing a large amount
of entropy and of the lightest superpartners, so the cosmological history of the
universe is generically non-thermal if our world is described by a
compactified M/string theory \cite{0804.0863,1006.3272}.

Thus the LHC Higgs boson looks like a fundamental particle, and looks like
data from a compactified constrained string theory with stabilized moduli
should look. We also learn that string theory is maturing into a useful
predictive framework that is relating explanations and tests, and that
M-theory compactified on a manifold of $G_{2}$ holonomy looks like a good
candidate for our string vacuum. The only scale in the compactified theory
is the Planck scale. Gaugino condensation gives supersymmetry breaking and
moduli stabilization, and the gravitino mass is calculated to be of order
50 TeV, solving the hierarchy problem. The resulting theory generically
has large trilinear couplings so the conditions for EWSB are generically
satisfied, with all the quantities in equation (1) for $M_Z$ of order 1-2 TeV. \
There is still a little hierarchy issue, but even with the gravitino, squarks
and Higgs mass matrix parameters of order 50 TeV, the little hierarchy
problem is not worse than in other models and theories, better than most. \
People who want to speak of 125 Gev as ``unnatural" should think about
whether they want to call results that emerge from a full theory
``unnatural". 

\acknowledgments{We have benefited from extensive interactions with Bobby Acharya and
Piyush Kumar, and we thank Aaron Pierce for recent helpful
discussions. The work of G.K., R.L, and B.Z. is supported by the DoE Grant DE-FG-02-95ER40899 and
and by the MCTP. R.L. is also supported by the String Vacuum Project Grant funded
through NSF grant PHY/0917807}

\bibliographystyle{h-physrev}
\bibliography{higgs}

\end{document}